\documentclass{article}

\usepackage{graphicx}
  \graphicspath{ {figures/} }
  \setkeys{Gin}{draft=false}
\usepackage{authblk}
\usepackage{natbib}

\newcommand{\pasp}{pp.}

\begin{document}

\title{Inverting OII 83.4 nm Dayglow Profiles Using Markov Chain Radiative Transfer}
\author[1]{George Geddes}
\author[2]{Ewan Douglas}
\author[1]{Susanna C. Finn}
\author[1]{Timothy Cook}
\author[1]{Supriya Chakrabarti}

\affil[1]{Lowell Center for Space Science and Technology,
University of Massachusetts, Lowell, Massachusetts, USA.}

\affil[2]{Department of Aeronautics and Astronautics, Massachusetts Institute of Technology,
Cambridge, Massachusetts, USA.}

\date{} 

\maketitle

\begin{abstract}
Emission profiles of the resonantly scattered OII~83.4~nm triplet can in principle be used to estimate \(\mathrm{O}^+\) density profiles in the F2 region of the ionosphere.
Given the emission source profile, solution of this inverse problem is possible, but requires significant computation. 
The traditional Feautrier solution to the radiative transfer problem requires many iterations to converge, making it time-consuming to compute. 
A Markov chain approach to the problem produces similar results by directly constructing a matrix that maps the source emission rate to an effective emission rate which includes scattering to all orders.
The Markov chain approach presented here yields faster results and therefore can be used to perform the \(\mathrm{O}^+\) density retrieval with higher resolution than would otherwise be possible.  
\end{abstract}

\section{Introduction}

One of the brightest features in the extreme ultraviolet (EUV) dayglow is the OII~83.4~nm  emission which is resonantly scattered by \(\mathrm{O}^+\)  ions. 
The resonant scattering increases the brightness and distribution of the 83.4~nm limb, and can be used to determine \(\mathrm{O}^+\) density in the ionosphere~\citep{kumar1983,mccoy1985,stephan2016}. 

The OII 83.4 nm  limb has been observed by the EUV spectrograph of the \textit{Remote Atmospheric and Ionospheric Detection System} (RAIDS EUVS), mounted on the International Space Station.
Past studies of limb profiles from RAIDS EUVS have shown good agreement with a radiative transfer model~\citep{douglas2012}.
\citet{picone1997} describe how to retrieve \(\mathrm{O}^+\) density from OII~83.4~nm limb intensities using Discrete Inverse Theory (DIT), but high-precision measurements to test the theory have been scarce. 
A new generation of ultraviolet airglow instruments are beginning operation in the next year. 
New missions such as the {\em Limb-Imaging Ionospheric and Thermospheric EUV Spectrograph} (LITES) ~\citep{LITES-SPIE}
 and the {\em Ionospheric Connection Explorer} (ICON) \citep{ICON-IEEE}
 will measure OII~83.4~nm dayglow with unprecedented precision.
 The ability to efficiently solve the inverse problem of \(\mathrm{O}^+\) density retrieval is advantageous to make full use of these data.
A real-time retrieval of ionospheric parameters from space-based airglow observations could constrain assimilative ionosphere models in regions where ground-based measurements are not possible, for instance.
Solving the inverse problem entails solving the OII~83.4~nm radiative transport problem for the  OII~83.4~nm emission (``the forward model'') many times.
The forward model used in the inversion procedure is critical to the physical validity of the result and the efficiency of the computation. %
The radiative transport for this application is typically calculated using an iterative method such as the Feautrier approximation~\citep{feautrier1964}.
However, the radiative transport problem can be formulated as a random walk which can be solved using a Markov chain with no iteration required~\citep{esposito1978}.
In this paper we will develop a Markov chain model for OII~83.4~nm resonant scattering, building on a model presented by~\citet{vickers1996}, compare it to an iterative method, and present an example of an \(\mathrm{O}^+\) density retrieval.
In section~\ref{section:scattering-model} we describe the Markov chain scattering model. 
In section~\ref{section:comparison} we compare an implementation of the Markov chain model to results from \textsf{AURIC}~\citep{AURIC}. 
In section~\ref{section:inversion} we apply a Bayesian estimation technique to retrieve an \(\mathrm{O}^+\) density profile from RAIDS EUVS OII~83.4~nm limb observations.

\section{Forward Model}
\label{section:scattering-model}

The radiative transport equation can be expressed as an integral equation which is often difficult to solve numerically. %
A common approach to its solution originates from \citet{feautrier1964} and is centered around a change of variables based on the two-stream approximation. %

\begin{figure}
  \centering
  \noindent\includegraphics[height=0.5\textwidth]{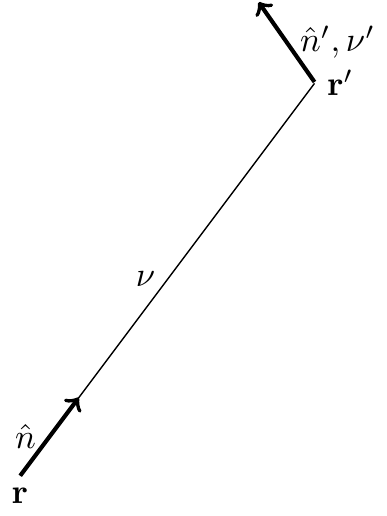}
  \caption{A photon is emitted at a location \(\mathbf r\) along the
    direction of the unit vector \(\hat n\) and at frequency
    \(\nu\). After scattering at \(\mathbf r'\), it carries on in some
    new direction \(\hat n'\) and frequency \(\nu'\).}
  \label{figure:step-diagram}
\end{figure}

Photons in the OII~83.4~nm triplet may be resonantly scattered by ground state \(\mathrm{O}^+\) ions as they travel through the ionosphere. 
The path taken by a photon in a scattering medium is a random walk and a Markov chain is a matrix containing the probabilities for each possible step~\citep{esposito1978}.
A photon's state is described by its direction of travel \((\hat n)\), its frequency \((\nu)\), and the location of the photon \((\mathbf r)\).  We will ignore polarization effects.
A step in the Markov chain is a transition from one state \((\mathbf r,\hat n,\nu)\) to another \((\mathbf r',\hat n',\nu')\), illustrated in Figure~\ref{figure:step-diagram}.
Given a photon in one state, it will make a transition to another state with some probability, \(P\) (Equation 1).
\(P\) is the product of \(P_T\), the transmission probability from \(\mathbf r\) to \(\mathbf r'\) and \(P_S\), the probability to scatter at the end of the transmission path.
The probabilities corresponding to each possible step from one state to another are the elements of the Markov chain matrix. %

\begin{equation}
  \label{step-prob}
  P\left(\mathbf r', \hat n', \nu' | \mathbf r, \hat n, \nu \right) = 
  P_T(\mathbf r'| \mathbf r,\hat n, \nu) P_S(\hat n',\nu'|\mathbf r', \hat n, \nu).
\end{equation}

The photon originating from \(\mathbf r\) will be absorbed or scattered at \(\mathbf r'\) with probability 

\begin{equation}
  P_T = e^{-\tau(\mathbf r,\mathbf r',\nu)},
\end{equation}

\noindent{}where \(\tau(\mathbf r,\mathbf r',\nu)\) is the optical depth along the path (Equation~\ref{optical-depth}), \(\hat n\) is a unit vector along the path (Equation~\ref{unit-vector}), and where \(\kappa(\mathbf r,\nu)\) is the extinction coefficient (Equation~\ref{ext-coeff}),

\begin{equation}
  \tau(\mathbf r,\mathbf r',\nu) = \int_0^{\left|\mathbf r' - \mathbf r\right|} \kappa(\mathbf r + \hat n s,\nu) \, d s
  \label{optical-depth}
\end{equation}
\begin{equation}
  \hat n = \frac{\mathbf{r}' - \mathbf{r}}{\left|\mathbf r' - \mathbf r\right|}
  \label{unit-vector}
\end{equation}
\begin{equation}
  \kappa(\mathbf r,\nu) = \sum_{\mathrm{species X}}
  [\mathrm{X}](z)\sigma_\mathrm{X}(\nu)
  \label{ext-coeff}
\end{equation}

The probability that the event that occurs at \(\mathbf r'\) is scattering rather than absorption is the albedo, \(\varpi(\mathbf r',\nu) = \frac{\kappa_{s}(\mathbf r', \nu)}{\kappa_{s}(\mathbf r', \nu) + \kappa(\mathbf r', \nu)}\).
The scattering probability \(P_s\) is the product of the albedo and some redistribution function, \(R(\hat n, \hat n', \nu, \nu', \mathbf r')\).
In general, the redistribution function can couple almost all of the variables together, making the integration difficult.
In the following sections, we will simplify the redistribution function by introducing the assumptions of complete frequency redistribution and a plane-parallel, azimuthally symmetric geometry.

\subsection{Complete Frequency Redistribution}
\label{section:CFR}
\label{section:lineshape}

Line center optical depths for the OII~83.4~nm emission in the terrestrial ionosphere are typically of order 1-10. %
This is  thick enough for scattering to change the emission intensity but small enough to assume the scattering is incoherent. %
In this case we make the approximation of complete frequency redistribution,

\begin{equation}
  R = p(\hat n, \hat n') \frac{\phi(a,\nu)}{\sqrt{\pi}\Delta\nu_D}  \frac{\phi(a,\nu')}{\sqrt{\pi}\Delta\nu_D'}, 
\end{equation}

\noindent{}where \(p\) can be chosen as either the Rayleigh phase function or \(\frac{1}{4\pi}\) for isotropic scattering, \(\Delta\nu_D = \sqrt{\frac{8kT\ln 2}{m_{\mathrm{O}}c^2}}\nu_0\) is the Doppler width, and \(\phi\) is the Voigt lineshape function \citep{meier1991}.

The lineshape of the scattering cross section describes how photons which are slightly off of the line center due to Doppler shift will interact with the scattering medium. 
The Voigt function is the convolution of the natural lineshape and a Doppler-broadened Gaussian lineshape. The unnormalized Voigt function is, 

\begin{equation}
  \label{voigt}
  \phi(a,x) = \frac{a}{\pi}\int_{-\infty}^\infty\frac{\exp(-y^2)}{(x-y)^2 + a^2}dy,
\end{equation}

\noindent{}where \(a = \frac{A}{4\pi\Delta\nu_D}\) is the Voigt parameter, \(A\) is the reciprocal of the natural lifetime and \(x = \frac{\nu-\nu_0}{\Delta\nu_D}\)  is a frequency parameter.%

\subsection{Geometry}
\label{section:geometry}

In order to cast the problem into a matrix form, we must divide the continuous space of photon states into discrete bins. 
For the spatial variables \(\mathbf r, \mathbf r'\), we chose an azimuthally symmetric, plane-parallel geometry. 
This choice simplifies the numerical implementation of the problem by allowing us to substitute the altitude variable \(z\) for the optical depth from infinity, 

\begin{equation}
  \label{tau-of-z}
  \tau(z,\nu) =  \int^\infty_z n(z)\sigma(T) \phi(a,\nu) dz + \sum_{\mathrm{x}} \int^\infty_z\kappa_\mathrm{x}(z) dz,
\end{equation}

\noindent{}where \(T\) is the temperature at altitude \(z\)  and \(\kappa_\mathrm{x}\) is the extinction cross section for neutral species x.

\begin{figure}
  \centering
  \noindent\includegraphics[width=0.5\textwidth]{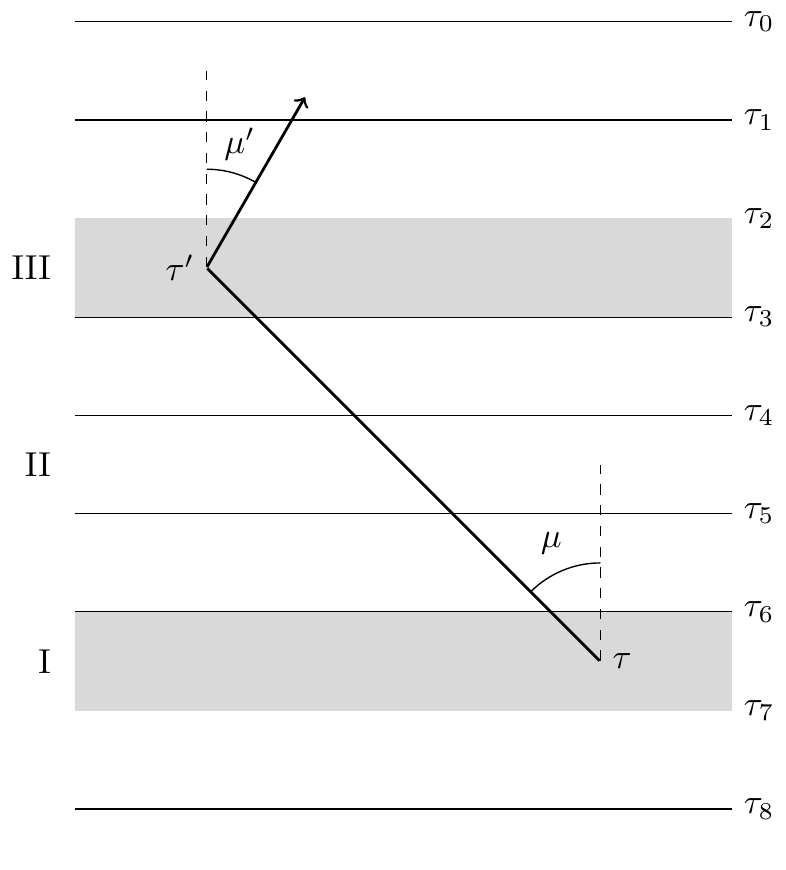}
  \caption{A photon emitted in region I at optical depth \(\tau\) at
    an angle \(\mu\) to the zenith escapes region I, passes through
    region II and is scattered into a new trajectory at angle \(\mu'\)
    at optical depth \(\tau'\) in region III.}
  \label{figure:tau-grid}
\end{figure}

We replace the continuous variable \(\tau\) with a discrete set of values \(\{\tau_1,\tau_2...,\tau_n\}\), with \(\tau_{i+1} > \tau_{i}\) and \(\tau_0 = 0 \), as shown in Figure~\ref{figure:tau-grid}. 

We will also consider a discrete set of zenith angles \(\{\mu_1,\mu_2...,\mu_m\}\)  at which photons can travel. We employ Legendre-Gauss quadrature to replace the integral over angle with a weighted sum,

\begin{equation}\label{legendre-gauss}
  \int_{-1}^{+1} f(\mu) d\mu \approx \sum_{i=1}^mw_if(\mu_i).
\end{equation}

With the assumptions made thus far, the mean probability of a single scattering event is, 

\begin{equation}
  \label{equation:nasty-thing}
  P_{ij} = \frac{
     \int_{\tau_{i}}^{\tau_i+1} d\tau \int_{\tau_j}^{\tau_{j+1}} d\tau' ww' e^{\left(\tau-\tau_{i}\right)/\mu} e^{\left(\tau_{i} - \tau_{j+1}\right)/\mu} e^{\left({\tau_{j+1}}-{\tau'}\right)/\mu} 
    \varpi(\tau',\nu)
    R(\mu,\mu',\nu,\nu',\tau)
  }
  {\int d\tau \int d\tau'},
\end{equation}

where \(\varpi = \frac{[\mathrm{O}^+]\sigma_{\mathrm{O}^+}}{\sum_{\mathrm{x}}[\mathrm{x}]\sigma_{\mathrm{x}}}\) is the single-scattering albedo. 
In order to integrate equation~(\ref{equation:nasty-thing}), we will assume that the temperature is nearly constant within each optical depth bin. 
We will also assume \(\varpi(\tau)\) is constant across each bin.
This will allow us to obtain an analytical expression for the average probability contained in a bin.
In practice, simply taking the albedo value from the top or bottom of the bin has little effect on model output.

To evaluate (\ref{equation:nasty-thing}) under these assumptions we use the integral, 

\begin{equation}
  \int_0^{\Delta\tau} e^{\tau} d\tau = (e^{\Delta\tau} - 1).
\end{equation}

The step probability averaged over a set of optical depth bins becomes,

\begin{equation}\label{equation:simplified-step}
  P_{ij} = \frac{ww'}{4}\varpi_jR \frac{\mu^2}{\Delta\tau_i\Delta\tau_j}(e^{-\Delta\tau_i/\mu}-1)(1-e^{-\Delta\tau_j/\mu})e^{-(\tau_j - \tau_i)/\mu},
\end{equation}

\noindent{}where \(\Delta\tau_i = \tau_{i+1} - \tau_i\), \(\varpi_j = \varpi(\tau_j)\), \(R = R(\mu,\mu',\nu,\nu',\tau)\), and \(w\) is the Gauss-Legendre weight corresponding to the angle \(\mu\).

\subsection{Markov Chain Formalism}

Now we may construct the single scattering matrix, \(Q\). We integrate the Equation~(\ref{equation:simplified-step}) over the frequency and angle variables to obtain the probability for a photon emitted at optical depth \(\tau_i\) to be scattered at \(\tau_j\),

\begin{equation}
  \label{single-scattering}
  Q_{ij} =  \sum_{a,b} \frac{w_a}{2} \frac{w_b}{2}
  \int_{-\infty}^\infty d\nu \int_{-\infty}^\infty d\nu'
  P_{ij}(\mu_a,\mu_b,\nu,\nu')
\end{equation}

The multiply scattered profile consists of photons scattered zero, one, two, three times, and so on in a geometric progression \(G = 1 + Q + Q^2 + \cdots = (1 - Q)^{-1}\). %
The final multiply-scattered profile, \(P_f\), is simply the product of the initial source distribution and the multiple-scattering matrix,

\begin{equation}
  P_f = P_0 (1-Q)^{-1}.
\end{equation}

The observed intensity profile can be obtained from the multiply scattered profile via another matrix, \(V\), which integrates the profile along sight lines. The intensity of the limb predicted by the model can be written,

\begin{equation}
  \hat y = V P_f.
\end{equation}

This can be rearranged to reveal that \(\hat y\) is the solution to a linear system of equations which is easier to compute than a matrix inversion:%

\begin{equation}
  \hat y(1-Q) = V P_0.
\end{equation}

This can be quite fast to calculate as long as the spatial bins are small enough that the errors incurred by averaging over the bin are not significant.

The plane-parallel assumption greatly simplifies the Feautrier solution.
It is possible to extend this solution to non-plane-parallel atmospheres, but at a high computational cost~\citep{anderson1977}.
In multiple dimensions, optical depth no longer increases monotonically with physical location, and therefore cannot be substituted for the location variables directly.

The Markov chain method described in this paper can accommodate multidimensional atmospheres. 
The formula for the matrix elements does not assume a particular spatial relationship between them.
The single scattering matrix can be calculated on any finite set of grid points.
The main change to the formula would be that the optical depth must be computed between cells as \(\tau_{ij}\) and not simply the difference of tabulated depths from the top of the atmosphere, \(\tau_j - \tau_i\). 

\subsection{Validation}
\label{section:comparison}

To validate our Markov chain radiative transfer approach model described above, we compare the results to a iterative  Feautrier solution. The radiative transfer equation with resonant scattering and frequency redistribution can be written, 

\begin{equation}
  \label{rt-equation}
  \mu\frac{dI}{d\tau}(\tau,\mu,\nu) = I(\tau,\mu,\nu) - S(\tau,\mu,\nu) - \int \,d\mu'\,d\nu'\, I(\tau,\mu',\nu') \varpi(\tau,\nu') R(\mu,\mu',\nu,\nu',\tau),
\end{equation}

\noindent{}where \(S(\tau,\mu,\nu)\) is the intensity of the initial photoemission and the integral expression represents the scattered component of the source function. The two terms can be combined into an effective source function which depends on the intensity,

\begin{equation}
  \label{effective-source}
  \mathcal{S}(I,\tau,\mu,\nu) = S(\tau,\mu,\nu) + \int \,d\mu'\,d\nu'\, I(\tau,\mu',\nu') \varpi(\tau,\nu') R(\mu,\mu',\nu,\nu',\tau).
\end{equation}

Next, we define a new set of variables,

\begin{equation}
  \label{flux-like}
  h(\tau,\mu,\nu) = \frac{1}{2}\left[I(\tau,\mu,\nu) - I(\tau,-\mu,\nu)\right],
\end{equation}

\begin{equation}
  \label{intensity-like}
  j(\tau,\mu,\nu) = \frac{1}{2}\left[I(\tau,\mu,\nu) + I(\tau,-\mu,\nu)\right].
\end{equation}

We can manipulate Equation~\ref{rt-equation} using these variables to obtain,

\begin{equation}
  \label{feautrier-equation}
  \mu^2\frac{d^2j}{d\tau^2}(\tau,\mu,\nu) = j(\tau,\mu,\nu) - \mathcal{S}(I,\tau,\mu,\nu).
\end{equation}

Equation~\ref{feautrier-equation} can be solved for \(j\) using a finite difference method, then we can compute \(I\) %
and use it to estimate \(\mathcal{S}\). Repeating this process will eventually converge to a solution.

\begin{figure} %
  \noindent\includegraphics[width=\textwidth]{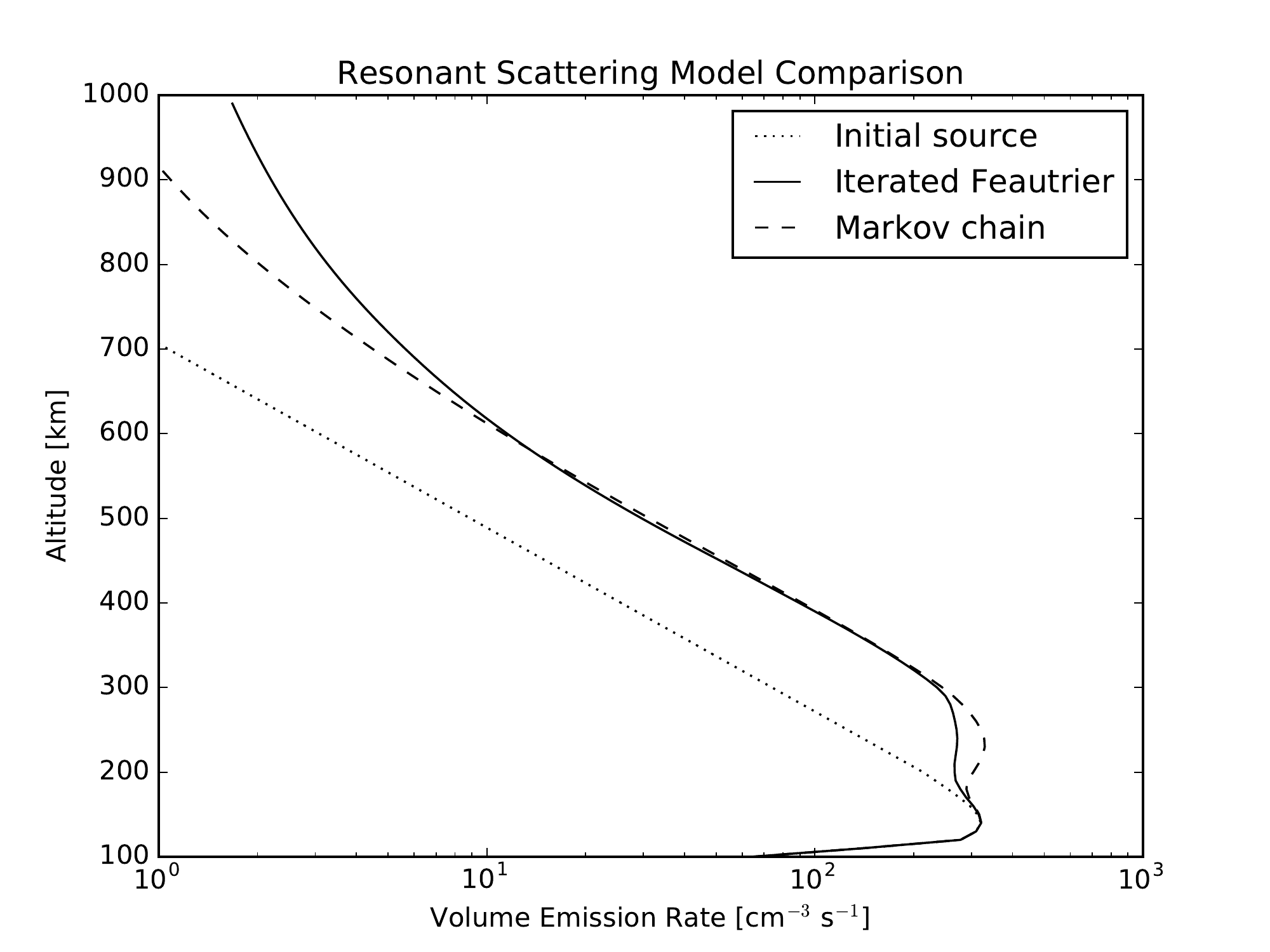}
  \caption{The Markov chain model agrees with the iterative Feautrier
    method very closely below 600 km. There is a discrepancy at the
    peak where optical depth and albedo are both changing
    rapidly. Inputs to the model are an MSISE-00 atmosphere and an
    IRI90 ionosphere with space weather conditions shown in
    Table~\ref{table:conditions}.}
  \label{figure:altitude-comparison}
\end{figure}

Figure~\ref{figure:altitude-comparison} shows agreement at the 10\% level between the two models below 500 km.
Above 500 km there is an excess of scattered light which grows to almost a factor of 2 difference from the Feautrier model at 1000 km.
However, this is unlikely to impact the retrieval for an instrument such as LITES, which will fly at 400 km and will not observe high altitudes.
Instruments that observe at higher altitudes should be able to distinguish between the two model predictions.

\section{Inverse Model}
\label{section:inversion}

In order to minimize overfitting, we chose a simple constant scale height Chapman\(-\alpha\) function to model the \(\mathrm{O}^+\) density profile,

\begin{equation}
  \label{equation:chapman} 
  n(z|N_m,h_m,H) = N_m\sqrt{\exp\left[1-\frac{z-h_m}{H}-\exp\left(-\frac{z-h_m}{H}\right)\right]},
\end{equation}

\noindent{}where \(N_m\) is the density at the peak of the ionosphere, \(h_m\) is the height of the peak of the ionosphere, and \(H\) is the scale height. This is a useful starting point, as it has been shown that this simple model of the ionosphere sufficiently approximates the height of the F2 peak for OII~83.4~nm modeling~\citep{douglas2012}. Models with additional parameters can yield good retrieval results as long as the parameters are nondegenerate, and additional parameterizations will be the topic of future work. 

In order to use the Markov chain forward model of section~\ref{section:scattering-model} we must supply an initial source of 83.4 nm photons.
The primary source of these is from excited \(\mathrm{O}^+\) ions decaying to the ground state.
Ions are excited into this state by solar photoionization, secondary electron impact ionization, and dissociative ionization of \(\mathrm{O}_2\).
There is also a negligible contribution from scattered solar photons~\citep{meier1991}.
The combined contribution of scattered solar flux, secondary electrons, and dissociative ionization is less than 10 percent of the total emission source, so we will consider only the \(\mathrm{O}^+(^4S)\leftarrow\mathrm{O}^+(^4P)\) photoemission rate in our initial source contribution~\citep{vickers1996}.
We compute the photoemission rate using the \emph{Atmospheric Ultraviolet Irradiance Code} (\textsf{AURIC})~\citep{AURIC} to simulate a scattered limb intensity profile using the forward model described in section~\ref{section:scattering-model}. 
\textsf{AURIC} employs NRLMSISE-90 and a modified version of the Hinteregger solar flux model to compute the initial photoproduction rate of OII~83.4~nm and other airglow features.
\citet{emmert2010} found that density of O during the 2007-2009 solar minimum was as much as 12\% below values predicted by MSIS, while other species were as much as 3\% below expected values. 
Errors in the modeled neutral atmosphere will of course affect the initial photoproduction rate as well. %
We denote the simulated limb intensity profile corresponding to the chapman parameters \(\{N_m,h_m,H\}\) as \(\hat{y}(\theta | N_m,H,h_m)\).

The limb intensity data from {RAIDS EUVS} are a vector of intensities $\mathbf y$ and look directions \(\mathbf \theta\).
The RAIDS EUVS scans across the limb to collect data with the tangent point altitudes ranging from 100~km to 300~km.
The data have been co-added over a 3 minute period and collected into 11 altitude bins to increase the signal to noise ratio (SNR).
The integration time in each altitude bin is therefore  180 s/11 \(\approx\) 16 s. 
The OII~83.4~nm emission is of order 500 Rayleighs, giving a SNR of \(\sqrt{16\times0.5\times500} \approx 63\), so we assume measurement errors follow a Gaussian distribution. %
Given a proposed set of Chapman parameters $\{N_m,H,h_m\}$ we compute the likelihood that the observed data were obtained from the corresponding emission predicted by our forward model
\citep{vanderplas_frequentism_2014},

\begin{equation}
  P(\mathbf y| N_m,H,h_m )=\prod_{i=1}^N \frac{1}{\sqrt{2\pi\sigma_i^2}}\exp{\left[-\frac{(y_i-\hat{y}(\theta_i | N_m,H,h_m))^2}{2\sigma_i^2}\right]}.
  \label{eq:likelihood}
\end{equation}

The retrieved parameters are those which maximize the likelihood of obtaining the observed data. 
In order to find the maximum likelihood parameters, we need to thoroughly sample the space of possible input parameters.
The {DIT} approach of \citet{picone1997} is a robust method to invert a physical system when uncertainties in model parameters and measurements are both approximately Gaussian. 
However, real observational uncertainties are not exactly Gaussian, particularly at high altitudes where the limb is not bright and photon counts are Poisson-distributed.
Model parameters retrieved from Gaussian observations will not have Gaussian errors since the parameters are simply linear functions of observed limb intensity.
In this case, {DIT} or the iterative methods used by \citet{vickers1996} may converge on a local minimum or fail to converge at all. 
Monte Carlo techniques randomly sample a parameter space in a way that is stable to anomalies and local minima \citep{sambridge_monte_2002}. 

Our model does propagate \(\mathrm{O}^+\) density parameters non-linearly, so we chose to sample the Chapman parameter space using a Markov chain Monte Carlo (MCMC) technique. 
Since each radiative transfer calculation is computationally intensive, a short algorithm convergence time minimizing the number of forward model runs is important for future efforts to parameterize the ionosphere in real time via 83.4 nm observation.
\citet{goodman_ensemble_2010} described an affine-invariant sampler which converges more rapidly than the more common Metropolis-Hastings {MCMC} method for skewed datasets where the likelihood is asymmetrical in parameter space which is used by the python package \textsf{emcee}~\citep{emcee}.
We used \textsf{emcee} to sample the Chapman parameter space for the retrieval of \(O^+\) density.
While faster convergence could be expected with a prior based on solar conditions or ground-based measurements, we begin with the na\"ive case of a uniform prior of Chapman parameters in the region \([20<H<60,200<h<300,7<\log_{10}N<11]\). 

\section{Results}
\label{section:results}
\begin{figure}
  \noindent\includegraphics[width=35pc]{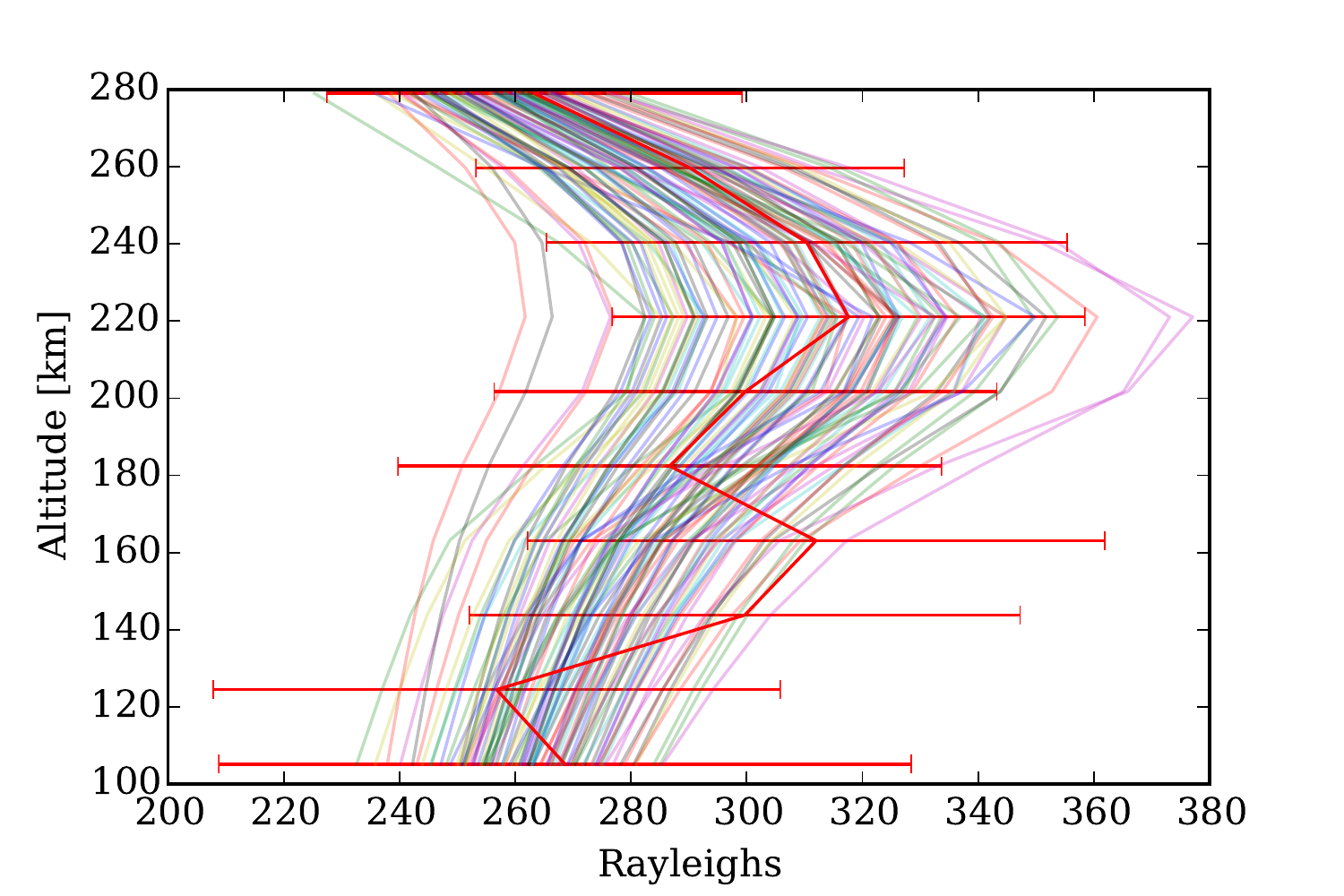}
  \caption{Comparison of the March 10, 2010 limb profile with 100 randomly drawn forward models after burn-in of the {MCMC} sampler. }
  \label{fig:10mar_100random}
\end{figure}

\begin{figure}
  \noindent\includegraphics[width=35pc]{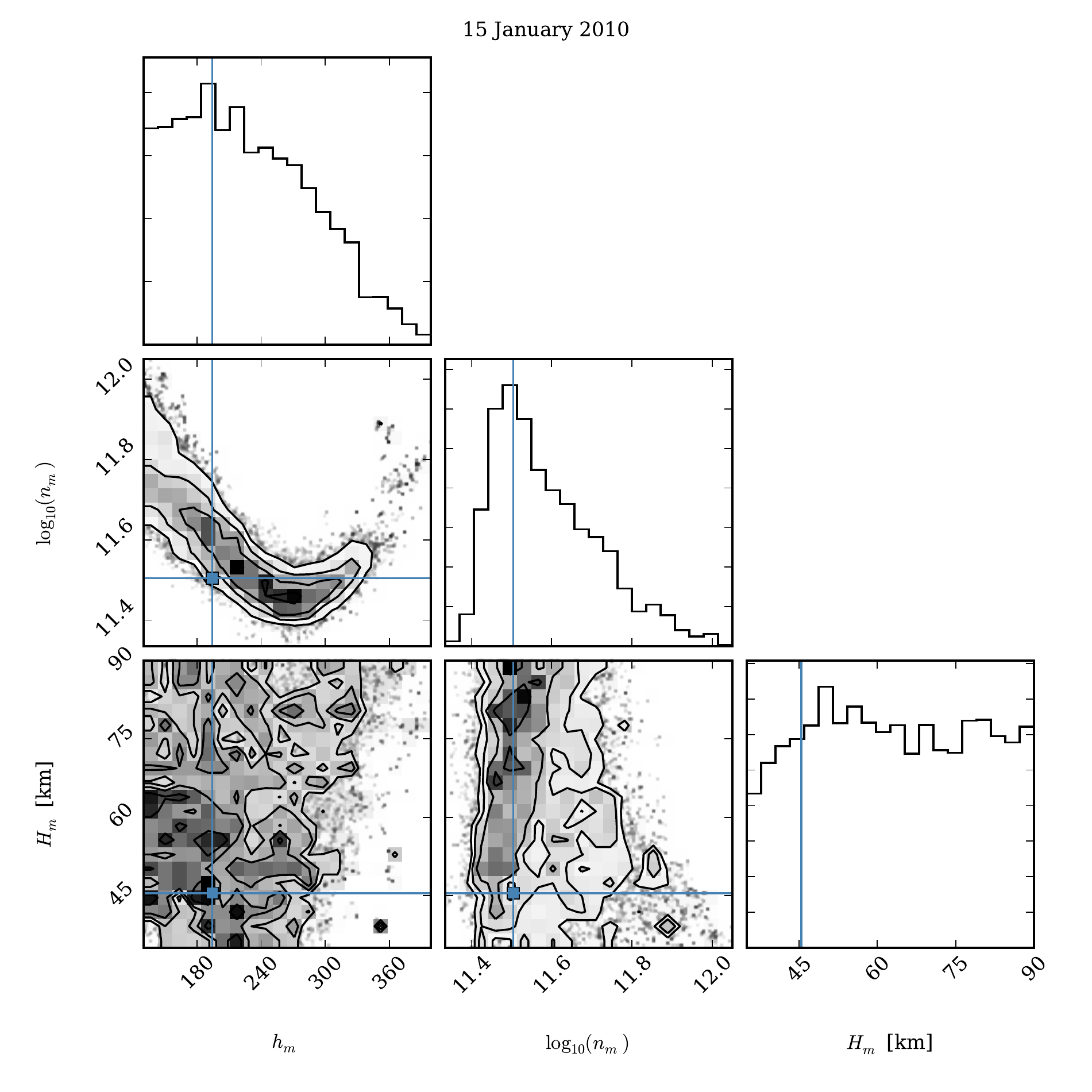}
  \caption{Posterior probability distributions for a three-parameter Chapman-$\alpha$ ionosphere given the January 15, 2010 {RAIDS} {EUVS} profile illustrating the morphological information content of 83.4 nm emission profiles. 
    Distributions were generated by $emcee$ sampling of parameter space and forward modeling with the Markov chain model.
    Ground truth Millstone Hill {ISR} best-fit Chapman-$\alpha$ parameters are shown as vertical lines in each histogram. 
    The source function was rescaled by a constant multiplicative calibration factor until the plasma density approximated the ground truth ISR best-fit value from \cite{douglas_evaluation_2012}. 
    (A rescaling factor of  1.7$\times S_0$  was used to generate this figure).
    The plasma density vs. peak height relation is well constrained and the peak of the scale-height probability agrees well with the ground based measurement, but the uncertainties are large and the individual parameters are not uniquely determined.  }
  \label{fig:15janMCMC}
\end{figure}

\begin{figure}
  \noindent\includegraphics[width=35pc]{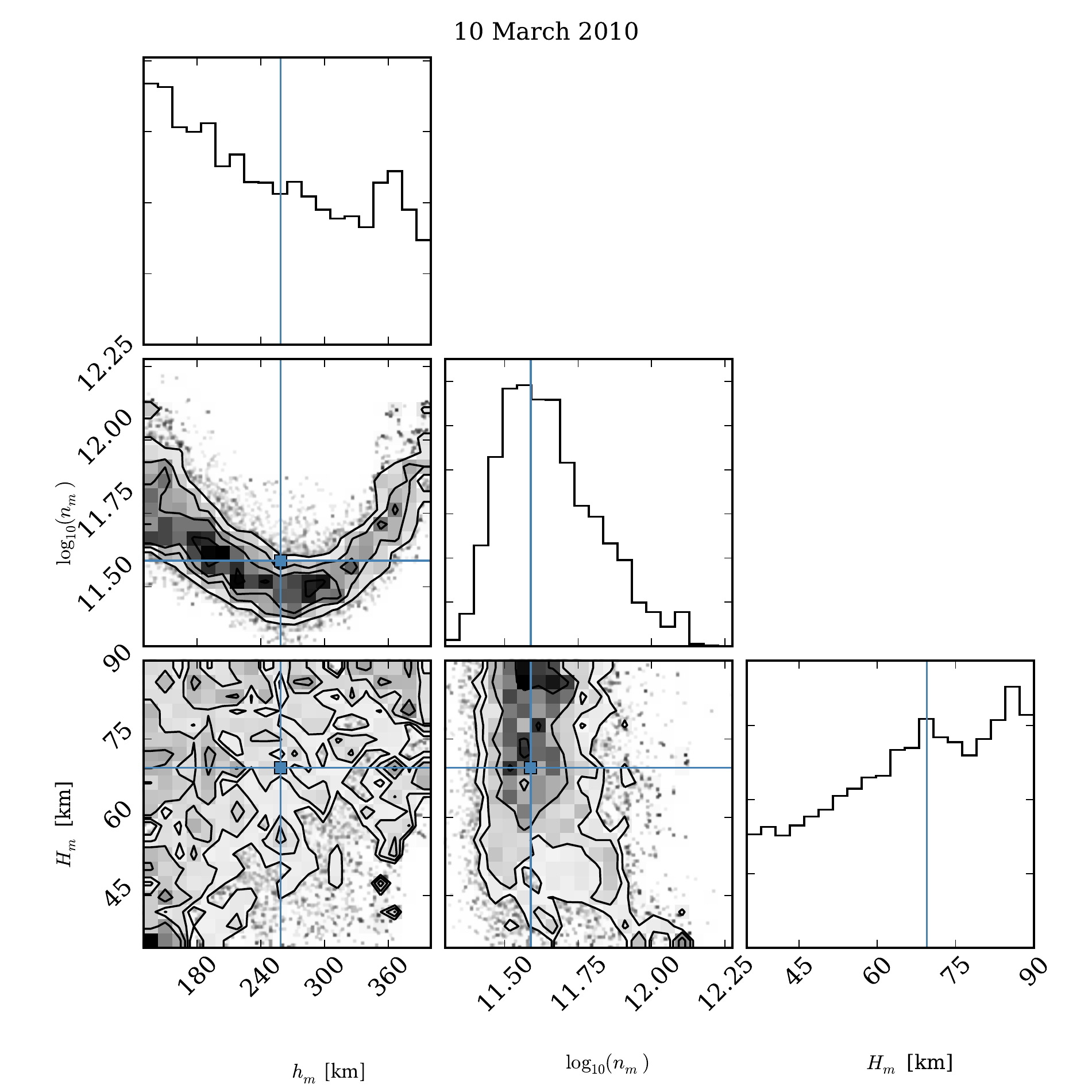}
  \caption{Posterior probability distributions for a three-parameter Chapman-$\alpha$ ionosphere given the March 10, 2010 {RAIDS} {EUVS} profile illustrating the morphological information content of 83.4 nm emission profiles. 
    Distributions were generated by $emcee$ sampling of parameter space and forward modeling with the Matrix Model.
    Ground truth Millstone Hill {ISR} best-fit Chapman-$\alpha$ parameters are shown as vertical lines in each histogram. 
  The initial source function is increased by a factor of 1.25, which is of the  order of the instrument calibration error, neglecting the source function and neutral density uncertainties.
    While a peak height versus density relation is still visible, the flatter emission profile shows increased degeneracy with the equiprobability region poorly localized compared to the January 15, 2010 example and lacks a single peak in the $H$ or \(h_m\) probabilities across expected values.}
  \label{fig:10marMCMC}
\end{figure}

We have tested our retrieval algorithm by applying it to RAIDS EUVS OII~83.4~nm limb data.
We compare the retrieval results to a coincident \(\mathrm{O}^+\) density measurement from the Millstone Hill incoherent scatter radar (ISR).
Although the ISR measures electron density rather than \(\mathrm{O}^+\) density, \(\mathrm{O}^+\) ions dominate the ionosphere between 200~km and 600~km, so there should be good agreement between the two.
In addition, the ISR measurement is associated with a fixed location while the RAIDS EUVS limb is integrated over about 1500~km of orbit, covering about 10 degrees of latitude.
The set of tangent points included in each of the two limb intensity profiles we use here is shown in Figure~1 of \citet{douglas2012}.
It is important to note that observed 83.4 nm limb intensity originates from a point closer to the observer than the tangent point of the line of sight, so information retrieved from such a measurement is about that point rather than the tangent point itself.
However the region with which the retrieval is best associated will depend on the retrieved parameters, so we will continue to associate retrievals with tangent point for convenience. 
The \textsf{emcee} configuration used for retrieval is shown in Table~\ref{table:emcee}. 
The configuration thoroughly sampled the parameter space with 26,400 forward models (48 walks of 550 steps each).
The total runtime was limited to 15 minutes due to system constraints.
Figure \ref{fig:10mar_100random} shows 100 randomly drawn forward models after the burn-in time, showing each realization is a qualitatively plausible fit  at the level of the forward models presented in \cite{douglas2012}.
Figures \ref{fig:15janMCMC} and \ref{fig:10marMCMC} show the posterior probability distributions for the same two observations as forward modeled from the Millstone Hill radar. 
In these ``corner'' plots \citep{foreman-mackey_triangle.py_2014} each histogram represents the marginal distribution of a single retrieved Chapman parameter.
Each two-dimensional plot represents the joint distribution of two Chapman parameters.
Although the prior distribution is uniform for each parameter, the retrieved (posterior) distributions are informed by observed data and measure confidence that a particular Chapman ionosphere could produce the observed scattered limb emission.
The ground-truth Millstone Hill {ISR} measurements of the corresponding Chapman$-\alpha$  parameters are shown as vertical lines in each histogram and intersecting lines in the density plots. 
Since there is a large uncertainty in the instrument calibration, %
the magnitude $S_0$ has been adjusted by an altitude constant scale factor such that peak of the retrieved \(N_m\) matches the ground-truth measurement.
(While similar and analogous to rescaling the final profile in the forward model comparison, adjusting $S_0$ rather than the final profile is biased toward correction of the volume excitation rate than the instrument calibration).
In the figures shown,  the multiplicative scaling factor for the 15 January, 2010 observation is 1.7$\times$ and on 1.25$\times$ 10 March, 2010.
While manual rescaling prohibits the single dataset retrieval of all three Chapman parameters without an external calibration, it shows that \(N_m\) can be retrieved given \(h_m\) or an external calibration.
Thus, this observation provides information constraining the ionosphere to a range of scale heights and peak altitudes.

While a degeneracy between peak height and peak density  can clearly be observed (first column, middle row) there is a well localized region of high probability slightly offset from the radar parameter peak for the 15 January 2010 dataset, while the scale-height (bottom row) is poorly constrained  but peaked near the true value.
There are only a small number of close overflights of RAIDS coincident with ISR measurements, so it is not possible to definitively determine whether the distinct offset between retrieved values and ground based measurements is due to differences in the ionospheric volume space sampled by the {ISR} and {RAIDS EUVS} observations or deficiencies in the forward model.
The more uniformly bright limb profile on 10 March 2010 provides less constraint on the probable scale height, implying the topside morphology (the decline in intensity at high altitudes visible on 15 Jan) drives the information content of the observation. 
An instrument which measures OII~83.4~nm emission at a wider range of tangent-point altitudes, either by observing from a higher altitude than the {ISS} \citep{mccoy1985} %
or by slowly rotating (e.g.  \cite{cotton_tomographic_2000}) would better constrain the morphology and narrow the retrieved parameter probabilities. 
Thus, another means of breaking the degeneracy between Chapman parameters is required.

\subsection{Other data sources} 
Previous studies of 83.4 nm emission have emphasized  the ``uniqueness'' of the ionospheric density  retrieved in the narrow sense of whether a single ``best-fit'' exists and closely approximates the expected ionosphere.
The best-fit approach breaks down for noisy data such as the presented RAIDS EUV profiles.
The results of the preceding {MCMC} retrieval analysis of RAIDS EUVS data show that additional information is required to constrain the retrieved ionosphere to the underlying physical parameters when observation uncertainty is too large. 
A Bayesian approach permits integrative analysis of the ionosphere, maximizing the leverage of even a poor measurement via straightforward inclusion of additional data. 
An auxiliary data source could improve the retrieval quality significantly given an appropriate prior distribution.
For example, a measurement that fixes either \(N_m\) or \(h_m\) would break the degeneracy between those parameters and allow the other to be retrieved accurately as well.
For example the peak density and peak height could be compared to results from the digisonde network of the Global Ionospheric Radio Observatory (GIRO).
Total electron content from GPS are available as well and would break the \(N_m,h_m\) degeneracy as well since they are proportional to the peak density but not the peak altitude. 

\subsection{Future Observations}
Retrievals informed by external data sources suffer from mismatches in calibration, thus there is also an incentive to increase the information recovered from a single EUV measurement.
The absolute calibration of the RAIDS EUVS is of order twenty-percent \citep{stephan_remote_2009,christensen_instrumentation_1992} while MSIS neutral oxygen densities have uncertainties exceeding twenty-percent. %
This uncertainty, coupled with uncertainty in the solar irradiance explains the difficulty in determining the source function intensity.
As presented previously, forward modeling of {RAIDS} data found a calibration error of up to 21\% \citep{douglas2012}. 

The RAIDS EUVS is mounted on a nodding platform with a  slit that has a \(2.4^\circ\) (horizontal) by \(0.1^\circ\) (vertical) field of view. This slit scans across the limb to sweep out a \(16.5^\circ\) field of regard. The soon-to-be-launched {LITES} employs a toroidal spectrograph and a \(10^\circ\times 10^\circ\) field of view to facilitate continuous spectral observation of multiple zenith angles \citep{cotton_single-element_1994}. 
The larger field of view enables LITES to observe all altitudes simultaneously, with a variable exposure time.  %
Furthermore, the LITES detector has a sensitivity of 15 counts per second per Rayleigh, 30 times higher than RAIDS EUVS~\citep{LITES-SPIE,RAIDS}.
Taking into account the different fields of view, the sensitivity along a particular look direction is 50\% higher in LITES than RAIDS EUVS, which will put tighter constraints on retrieved parameters.

The co-adding of RAIDS EUVS measurements to generate profiles requires including measurements over a wide geographic range. 
Thus, alternatively, the increased {LITES} signal allows binning observations to SNR equal to that of {RAIDS EUVS} for increased spatial resolution.
A {LITES} measurement of the same signal-to-noise as these RAIDS EUVS observation would cover a much shorter ground track and corresponding ionospheric volume, reducing the variability in measured emission rate due to small and medium scale structures such as traveling ionospheric disturbances and the equatorial ionization anomaly.

The increased precision will tightly constrain the morphology, suggesting value in attempting to retrieve alternate parameterizations of the \(\mathrm{O}^+\) density, such as a linearly changing scale height Chapman layer model.
There is still an uncertainty associated with the initial production of the 83.4 nm emission.
The production source will be measured indirectly by observing the optically thin OII~61.7~nm feature~\citep{stephan2012}.
OII~61.7~nm shares the same production mechanism as OII~83.4~nm, but it is not scattered at all.
This feature is not as bright as OII~83.4~nm, and the assumption of Gaussian observation uncertainty may not hold. 
Making use of this measurement in the Bayesian retrieval method is possible, but will require a model describing the exact ratio of the two emission rates.
Including an observation of OII~61.7~nm is expected to improve accuracy by eliminating the dependence on solar EUV flux and neutral density models, but further work is required to confirm and quantify this improvement.

\section{Conclusion}

We have developed a linearized solution to the radiative transfer problem. 
This solution is useful for retrieving \(\mathrm{O}^+\) density profiles from OII~83.4~nm limb observations. %

We have shown how to compute a matrix that transforms source emission profiles to multiply-scattered dayglow profiles in a plane-parallel atmosphere and ionosphere.
This method does not require iteration of the solution, and is therefore very straightforward to apply.
We have compared this method to the more commonly used Feautrier method, and have shown that it yields similar results.

We have also presented a Bayesian approach to retrieval of a Chapman layer \(\mathrm{O}^+\) density profile from OII~83.4~nm limb observations.

\section*{Acknowledgments}
  Computations for the Bayesian parameter retrieval were performed on
  the Boston University Scientific Computing Cluster, an interface to
  the Massachusetts Green Computing Facility.
  Initial OII~83.4~nm photoproduction rates were generated by AURIC,
  by Computational Physics Inc.  Code for the Markov chain model
  results was based on code written by John S.\ Vickers, and can be
  found at {https://github.com/georgegeddes/mcrt}.  This work is
  supported in part by Office of Naval Research Award \#14639, NSF
  Grant No.\ AGS-1315354, and a summer fellowship from the
  Massachusetts Space Grant Consortium.  RAIDS is integrated and flown
  as part of the HREP experiment under the direction of the Department
  of Defense Space Test Program. RAIDS was built jointly by the NRL
  and The Aerospace Corporation with additional support from the
  Office of Naval Research. 

\bibliographystyle{plainnat}

\begin{thebibliography}{26}
\providecommand{\natexlab}[1]{#1}
\expandafter\ifx\csname urlstyle\endcsname\relax
  \providecommand{\doi}[1]{doi:\discretionary{}{}{}#1}\else
  \providecommand{\doi}{doi:\discretionary{}{}{}\begingroup
  \urlstyle{rm}\Url}\fi

\bibitem[{\textit{Anderson and Hord}(1977)}]{anderson1977}
Anderson, D.~E., and C.~W. Hord (1977), Multidimensional radiative transfer:
  Applications to planetary coronae, \textit{Planetary and Space Science},
  \textit{25}(6), 563--571.

\bibitem[{\textit{Christensen et~al.}(1992{\natexlab{a}})\textit{Christensen,
  Kayser, Pranke, Straus, Gutierrez, Chakrabarti, McCoy, Meier, Wolfram, and
  Picone}}]{christensen_instrumentation_1992}
Christensen, A.~B., D.~C. Kayser, J.~B. Pranke, P.~R. Straus, D.~J. Gutierrez,
  S.~Chakrabarti, R.~P. McCoy, R.~R. Meier, K.~D. Wolfram, and J.~M. Picone
  (1992{\natexlab{a}}), Instrumentation on the {RAIDS} ({Remote} {Atmospheric}
  and {Ionospheric} {Detection} {System}) experiment. {II} - {Extreme}
  ultraviolet spectrometer, photometer, and near {IR} spectrometer,
  \textit{1745}.

\bibitem[{\textit{Christensen et~al.}(1992{\natexlab{b}})\textit{Christensen,
  Kayser, Pranke, Straus, Gutierrez, Chakrabarti, McCoy, Meier, Wolfram, and
  Picone}}]{RAIDS}
Christensen, A.~B., D.~C. Kayser, J.~B. Pranke, P.~R. Straus, D.~J. Gutierrez,
  S.~Chakrabarti, R.~P. McCoy, R.~R. Meier, K.~D. Wolfram, and J.~Picone
  (1992{\natexlab{b}}), Instrumentation on the raids experiment ii:
  extreme-ultraviolet spectrometer, photometer, and near-ir spectrometer, in
  \textit{San Diego'92}, pp. 89--98, International Society for Optics and
  Photonics.

\bibitem[{\textit{Cotton et~al.}(1994)\textit{Cotton, Cook, and
  Chakrabarti}}]{cotton_single-element_1994}
Cotton, D.~M., T.~Cook, and S.~Chakrabarti (1994), Single-element imaging
  spectrograph, \textit{Appl. Opt.}, \textit{33}(10), 1958--1962,
  \doi{10.1364/AO.33.001958}.

\bibitem[{\textit{Cotton et~al.}(2000)\textit{Cotton, Stephan, Cook, Vickers,
  Taylor, and Chakrabarti}}]{cotton_tomographic_2000}
Cotton, D.~M., A.~Stephan, T.~Cook, J.~Vickers, V.~Taylor, and S.~Chakrabarti
  (2000), Tomographic {Extreme}-{Ultraviolet} {Spectrographs}: {TESS},
  \textit{Appl. Opt.}, \textit{39}(22), 3991--3999, \doi{10.1364/AO.39.003991}.

\bibitem[{\textit{{Douglas} et~al.}(2012)\textit{{Douglas}, {Stephan},
  {Bishop}, {Budzien}, {Christensen}, {Hecht}, and
  {Chakrabarti}}}]{douglas2012}
{Douglas}, E.~S., A.~W. {Stephan}, R.~L. {Bishop}, S.~A. {Budzien}, A.~B.
  {Christensen}, J.~H. {Hecht}, and S.~{Chakrabarti} (2012), {Ionospheric
  Observations with Raids, AN Extensive Comparison of O$^{+}$ 83.4 NM Emission
  to Ground Based Observations}, \textit{AGU Fall Meeting Abstracts}.

\bibitem[{\textit{Douglas et~al.}(2012)\textit{Douglas, Smith, Stephan,
  Cashman, Bishop, Budzien, Christensen, Hecht, and
  Chakrabarti}}]{douglas_evaluation_2012}
Douglas, E.~S., S.~M. Smith, A.~W. Stephan, L.~Cashman, R.~L. Bishop, S.~A.
  Budzien, A.~B. Christensen, J.~H. Hecht, and S.~Chakrabarti (2012),
  Evaluation of ionospheric densities using coincident {OII} 83.4 nm airglow
  and the {Millstone} {Hill} {Radar}, \textit{J. Geophys. Res.}, \textit{117},
  8 PP., \doi{201210.1029/2012JA017574}.

\bibitem[{\textit{Emmert et~al.}(2010)\textit{Emmert, Lean, and
  Picone}}]{emmert2010}
Emmert, J., J.~Lean, and J.~Picone (2010), Record-low thermospheric density
  during the 2008 solar minimum, \textit{Geophysical Research Letters},
  \textit{37}(12).

\bibitem[{\textit{Esposito and House}(1978)}]{esposito1978}
Esposito, L., and L.~House (1978), Radiative transfer calculated from a markov
  chain formalism, \textit{The Astrophysical Journal}, \textit{219},
  1058--1067.

\bibitem[{\textit{{Feautrier}}(1964)}]{feautrier1964}
{Feautrier}, P. (1964), {A Procedure for computing the Mean Intensity and the
  Flux}, \textit{SAO Special Report}, \textit{167}, 80.

\bibitem[{\textit{{Foreman-Mackey} et~al.}(2013)\textit{{Foreman-Mackey},
  {Hogg}, {Lang}, and {Goodman}}}]{emcee}
{Foreman-Mackey}, D., D.~W. {Hogg}, D.~{Lang}, and J.~{Goodman} (2013), {emcee:
  The MCMC Hammer}, \textit{\pasp}, \textit{125}, 306--312,
  \doi{10.1086/670067}.

\bibitem[{\textit{Foreman-Mackey et~al.}(2014)\textit{Foreman-Mackey,
  Price-Whelan, Ryan, {Emily}, Smith, Barbary, Hogg, and
  Brewer}}]{foreman-mackey_triangle.py_2014}
Foreman-Mackey, D., A.~Price-Whelan, G.~Ryan, {Emily}, M.~Smith, K.~Barbary,
  D.~W. Hogg, and B.~J. Brewer (2014), triangle.py v0.1.1.

\bibitem[{\textit{Goodman and Weare}(2010)}]{goodman_ensemble_2010}
Goodman, J., and J.~Weare (2010), Ensemble samplers with affine invariance,
  \textit{Communications in Applied Mathematics and Computational Science},
  \textit{5}(1), 65--80, \doi{10.2140/camcos.2010.5.65}.

\bibitem[{\textit{Kumar et~al.}(1983)\textit{Kumar, Chakrabarti, Paresce, and
  Bowyer}}]{kumar1983}
Kumar, S., S.~Chakrabarti, F.~Paresce, and S.~Bowyer (1983), The {O}+ 834-Å
  dayglow: {Satellite} observations and interpretation with a radiation
  transfer model, \textit{Journal of Geophysical Research: Space Physics},
  \textit{88}(A11), 9271--9279, \doi{10.1029/JA088iA11p09271}.

\bibitem[{\textit{McCoy et~al.}(1985)\textit{McCoy, Jr, and
  Chakrabarti}}]{mccoy1985}
McCoy, R.~P., D.~E.~A. Jr, and S.~Chakrabarti (1985), F 2 {Region} {Ion}
  {Densities} from {Analysis} of {O}+ 834-Å {Airglow}: {A} {Parametric}
  {Study} and {Comparisons} with {Satellite} {Data}, \textit{Journal of
  Geophysical Research}, \textit{90}(A12), PP. 12,257--12,264,
  \doi{198510.1029/JA090iA12p12257}.

\bibitem[{\textit{Meier}(1991)}]{meier1991}
Meier, R.~R. (1991), Ultraviolet spectroscopy and remote sensing of the upper
  atmosphere, \textit{Space Science Reviews}, \textit{58}, 1--185.

\bibitem[{\textit{Picone et~al.}(1997)\textit{Picone, Meier, Kelley, Dymond,
  Thomas, Melendez-Alvira, and McCoy}}]{picone1997}
Picone, J.~M., R.~R. Meier, O.~A. Kelley, K.~F. Dymond, R.~J. Thomas, D.~J.
  Melendez-Alvira, and R.~P. McCoy (1997), Investigation of ionospheric {O}+
  remote sensing using the 834-Å airglow, \textit{Journal of Geophysical
  Research}, \textit{102}, 2441--2456, \doi{10.1029/96JA03314}.

\bibitem[{\textit{Rider et~al.}(2015)\textit{Rider, Immel, Taylor, and
  Craig}}]{ICON-IEEE}
Rider, K., T.~Immel, E.~Taylor, and W.~Craig (2015), Icon: Where earth's
  weather meets space weather, in \textit{2015 IEEE Aerospace Conference}, pp.
  1--10, IEEE.

\bibitem[{\textit{Sambridge and Mosegaard}(2002)}]{sambridge_monte_2002}
Sambridge, M., and K.~Mosegaard (2002), Monte carlo methods in geophysical
  inverse problems, \textit{Reviews of Geophysics}, \textit{40}(3).

\bibitem[{\textit{Stephan}(2016)}]{stephan2016}
Stephan, A.~W. (2016), Advances in remote sensing of the daytime ionosphere
  with euv airglow, \textit{Journal of Geophysical Research: Space Physics}.

\bibitem[{\textit{Stephan et~al.}(2009)\textit{Stephan, Budzien, Bishop,
  Straus, Christensen, Hecht, and Van~Epps}}]{stephan_remote_2009}
Stephan, A.~W., S.~A. Budzien, R.~L. Bishop, P.~R. Straus, A.~B. Christensen,
  J.~H. Hecht, and Z.~Van~Epps (2009), The {Remote} {Atmospheric} and
  {Ionospheric} {Detection} {System} on the {ISS}: sensor performance and space
  weather applications from the extreme to the near ultraviolet, pp.
  74,380Y--74,380Y, \doi{10.1117/12.825167}.

\bibitem[{\textit{Stephan et~al.}(2012)\textit{Stephan, Picone, Budzien,
  Bishop, Christensen, and Hecht}}]{stephan2012}
Stephan, A.~W., J.~M. Picone, S.~A. Budzien, R.~L. Bishop, A.~B. Christensen,
  and J.~H. Hecht (2012), Measurement and application of the {O} {II} 61.7 nm
  dayglow, \textit{Journal of Geophysical Research: Space Physics},
  \textit{117}(A1), n/a--n/a, \doi{10.1029/2011JA016897}.

\bibitem[{\textit{{Stephan} et~al.}(2014)\textit{{Stephan}, {Budzien}, {Finn},
  {Cook}, {Chakrabarti}, {Powell}, and {Psiaki}}}]{LITES-SPIE}
{Stephan}, A.~W., S.~A. {Budzien}, S.~C. {Finn}, T.~A. {Cook},
  S.~{Chakrabarti}, S.~P. {Powell}, and M.~L. {Psiaki} (2014), Ionospheric
  imaging using merged ultraviolet airglow and radio occultation data, in
  \textit{Imaging Spectrometry XIX}, \textit{Proceedings of the SPIE}, vol.
  9222, p. 92220M, \doi{10.1117/12.2061420}.

\bibitem[{\textit{Strickland et~al.}(1999)\textit{Strickland, Bishop, Evans,
  Majeed, Shen, Cox, Link, and Huffman}}]{AURIC}
Strickland, D., J.~Bishop, J.~Evans, T.~Majeed, P.~Shen, R.~Cox, R.~Link, and
  R.~Huffman (1999), Atmospheric ultraviolet radiance integrated code (auric):
  Theory, software architecture, inputs, and selected results, \textit{Journal
  of Quantitative Spectroscopy and Radiative Transfer}, \textit{62}(6),
  689--742.

\bibitem[{\textit{VanderPlas}(2014)}]{vanderplas_frequentism_2014}
VanderPlas, J. (2014), Frequentism and {Bayesianism}: {A} {Python}-driven
  {Primer}, \textit{arXiv:1411.5018 [astro-ph]}, arXiv: 1411.5018.

\bibitem[{\textit{Vickers}(1996)}]{vickers1996}
Vickers, J.~S. (1996), An evaluation of {EUV} remote sensing of the ionosphere
  using the {OII} 834-\r{A} emission, Ph.D. thesis, {UC} Berkeley.

\end{thebibliography}

 %
 %
 %
 %
 %

\begin{table}
  \caption{Input parameters for Figure~\ref{figure:altitude-comparison}.}
  \begin{center}
    \begin{tabular}{c|c}
      Parameter & Value \\
      \hline
      Date & 2010-03-10\\
      UTC  & 15:00:00\\
      Geographic Latitude  & 42\\
      Geographic Longitude & 288\\
      F107 & 79.3\\
      F107 Average & 80.7\\
      ap & 9\\
    \end{tabular}
  \end{center}
  \label{table:conditions}
\end{table}

\begin{table}
  \caption{ Inputs used for the Monte Carlo sampler \textsf{emcee} in the retrieval tests.
    The walker number and step length were tuned to keep the runtime below 15 minutes while still achieving good coverage of parameter space.  Runtime is for a \textit{Intel(R) Xeon(R) CPU E5-2670 v2} at 2.50GHz with 10 cores and a maximum of 20 threads per core on the Massachusetts Green High Performance Computing Center.}%
  \begin{center}
    \begin{tabular}{c|c}
      Parameter  & Value\\
      \hline
      Dimensions &   3 \\
      Walkers    &  48 \\
      Steps      & 550 \\
      $n_{burn}$ &  50 \\
      Typical Runtime [min.] & 14.5 \\
      Threads    & 30 \\
      $H$ Initial Guess range [km] & 20-60  \\
      \(h_m\) Initial Guess range [km] & 200-300 \\
      \(N_m\)  initial Guess range [$m^3$] & $5\times10^{7}-5\times10^{11}$ \\
    \end{tabular}
  \end{center}
  \label{table:emcee}
\end{table}%
\end{document}